\documentstyle[12pt]{article}

\begin{document}
\title{Nuclear Suppression of Hadroproduction}
\author{S.V.Akulinichev\\
{\it Institute for Nuclear Research,}\\
{\it 60-th October Anniversary Prospect 7a, Moscow 117312, Russia.}}
\maketitle
\begin{abstract}
We demonstrate that the main part of
the observed nuclear suppression of dilepton and charmonium
production, especially the $x$-dependence of the suppression,
can be explained by the absorption of fast initial partons.
We assume the absorption to be $x_{1}$-dependent and determine the
empirical form of this dependence.
Several reactions at different energies
are described with the same parametrization of
the absorption cross section. The factorization theorem constraints
and some alternative models are discussed.
The obtained description of data allows to
make conclusions about the origin of nuclear effects in
hadro- and electroproduction. In particular,  it is concluded that
the parton recombination and the excess pion contribution are unimportant
in nuclei and an indication of excess gluons at $x_{2}\sim 0.1$ is
observed.

\end{abstract}
\newpage
{\bf I. INTRODUCTION}

Several effects have been considered in order to explain the
observed strong suppression of $J/\psi-$production
in hadron-nucleus collisions\cite{Badier,Alde1}.
We can mention some of them: the intrinsic charm\cite{Brod1}, the comover
interaction\cite{Brod2}, the recombination of nuclear partons
\cite{recomb1,recomb2}, the interaction of non-singlet final parton
configurations\cite{Weise1,recomb2}, the
elastic rescattering of initial partons\cite{Pirner},
the energy loss of initial partons\cite{Eloss}  and the absorption of initial
partons\cite{Akul1}. However, we agree with the authors of Ref.\cite{Hufner}
that the $x-$dependence of charmonium production suppression has not yet
been described.
Without a reliable description of the charm production suppression
in hadron-nucleus interactions, it is hard to analyze this effect in heavy
ion collisions while searching for the quark-gluon plasma signals\cite{Satz}.
Moreover, a nuclear suppression has been observed in the
Drell-Yan (DY) production by pions\cite{Bordalo} and, though much less,
by protons\cite{Alde2}. A self-consistent description of hadroproduction
suppression in nuclei, as well as of the nuclear
effects in electroproduction\cite{NMC}, is now one of the important goals of
high energy nuclear physics. We here develop a model, which successfully
describes  several reactions at different energies
and, as we try to argue, does not contradict the true factorization forecasts.

We now briefly overview the status of the above mentioned models.
1) The intrinsic
charm model is, in principle, able to account for the observed $x_{F}-$
dependence of the $J/\psi-$ suppression, namely the suppression increasing
with $x_{F}$. However, the experimental searches for the intrinsic charm
in protons\cite{intr} have not revealed any charm signals. With the
experimental
upper limit on the amount of the intrinsic charm in protons\cite{intr}, this
mechanism cannot explain the observed charm production suppression. Moreover,
this mechanism is irrelevant for the DY production by pion, where the
suppression is also significant, as it will be shown below.
2) The comover interaction
is most effective at negative $x_{F}$, because the density of comovers
increases for less energetic produced quarkonia. In fact, a strong
suppression was observed in quarkonium production at negative
$x_{F}$\cite{Alde3}.
In the present article we are concerned with the region of positive medium
and large $x_{F}$ of produced charmonia, where
the comover interaction must be minor.
3) In agreement with the strong factorization, the parton recombination
models suggest that the initial state
effects in hadroproduction are included in nuclear
structure functions. It has been assumed\cite{recomb1} that nuclear structure
functions are suppressed at small $x_{2}$ due to QCD recombination of partons
from neighbour nucleons. This mechanism has been widely used
to explain the nuclear shadowing in electroproduction at low $x_{2}$.
This  effect would lead to an $x_{2}-$scaled
nuclear suppression of quarkonium and dilepton hadroproduction. However,
a comparative analysis of the data at different proton energies
clearly shows\cite{Alde1,Hufner} that the $J/\psi-$suppression is scaled
on $x_{F}$ or $x_{1}$, but not on $x_{2}$. It seems that these data create
serious troubles for the nuclear parton recombination hypothesis, as we
will argue below.
4) Final state interactions of produced non-singlet
parton configurations can be responsible for a suppression of
quarkonium production, because these interactions are not restricted by the
color transparency.
However, there is no physical reason why this effect
should increase with $x_{F}$ at large $x_{F}$. Moreover, this effect
cannot explain
the DY production suppression. Therefore the final state interactions
of produced heavy quarks cannot be the only mechanism of nuclear
suppression. 5) The elastic rescattering of initial partons naturally explains
the dilepton and quarkonium yield increasing with $p_{T}$. However this
mechanism alone, if not completed by the energy loss of partons, makes no
predictions for
the $x-$dependence of hadroproduction suppression. 6) The two remaining
models, assuming the energy loss (EL) of fast initial partons due to soft
elastic rescattering\cite{Eloss} and the absorption
of those partons\cite{Akul1}, will be considered below in more detail.
We first discuss these two models with respect to the factorization
theorem.

The cross section factorization in large-$Q^{2}$ reactions can be represented,
in the case of DY production, in the following form\cite{CSS,Bodw}
\begin{equation}
d\sigma^{DY}\sim\int^{1}_{0}\frac{dx_{1}}{x_{1}}\frac{dx_{2}}{x_{2}}
f_{1}^{DY}(x_{1})\:H(x_{1},x_{2},Q^{2})\:f_{2}^{DY}(x_{2}),
\end{equation}
where $f^{DY}(x)$ are parton densities containing all  non-perturbative
long-distance effects and $H$, being a  projectile- and
target-independent function,
describes the hard parton-parton subprocess. The integration
over transverse momenta and the $Q^{2}-$evolution of structure
functions are implied in Eq.(1). In the lab. frame,
the subscripts ${1}$ or ${2}$ correspond to projectile or target, respectively.
In a general case, when the functions $f(x)$ depend on the reaction under
consideration, the form (1) corresponds to the weak factorization.
The strong factorization means that the structure functions are
reaction-independent, e.g. the structure functions
measured in electroproduction are equal to $f^{DY}(x)$  (integrated  over
transverse momenta). The limitations of the factorization have been
discussed in several papers (see, e.g., Refs.\cite{Bodw}-\cite{Brod3}).
In particular, it was obtained\cite{BBL,Bodw} that in the case of collinear
active-spectator parton interactions, followed by the productive annihilation
of the active parton, the strong factorization is valid if
\begin{equation}
Q^{2}\gg x_{2}L M l_{T}^{2},
\end{equation}
where $L$ and $M$ are the target length (in the rest frame) and mass and
$l_{T}$ is the transverse momentum, transferred in the active-spectator
interaction. From Eq.(2), it is reasonable to expect that the corrections
to factorization, in the case of
collinear active-spectator rescattering, should vanish as $1/Q^{2}$. It was
also concluded\cite{Brod3} that the maximum energy loss of an incident parton
due to induced gluon radiation, which is expected to be the main source
of energy losses, is
\begin{equation}
\Delta_{max} x_{1}=k_{T}^{2}L/2E_{lab},
\end{equation}
where $k_{T}$ is the transverse momentum of the radiated gluon and $E_{lab}$
is the projectile energy. This corresponds to an average parton energy loss
per unit length $dE/dz\sim 0.3 GeV/fm$\cite{Brod3}.
The limitations (2) and (3) directly concern the EL-model,
because this model assumes the initial state interactions of the type
studied in Refs.\cite{BBL,Bodw,Brod3}. This means that in the parametrization
for the average parton energy loss in nuclei\cite{Eloss},
\begin{equation}
\Delta x_{1} = \kappa x_{1} C_{i} (Q_{0}/Q)^{n}A^{1/3}
\end{equation}
($C_{g}=3$ for gluons, $C_{q}=4/3$ for partons and $Q_{0} \approx 4 GeV$),
it is more reasonable to use $n$=2. This choice is also supported by the
observation that the leading contribution to the DY production is twist 4.
The limitation (3) was significantly violated with $\kappa\approx$0.003,
used in Ref.\cite{Eloss} to fit the data for the charmonium suppression.
Nevertheless we will present below some  numerical results of the
EL-model, considering this model as a possible alternative for our model.

The absorption of fast initial partons by nuclear nucleons corresponds to
all parton-parton processes, which remove the active parton from the
production channel. These processes are not followed by the productive
annihilation of the active parton, in contrast to initial state interactions
assumed in the EL-model.   An example of such processes is a
particle production on front nucleons: though the formation of produced
particles takes a long time, the momentum transfer in productive
parton-parton collisions
can be quite large, of the order of the projectile momentum
$P$. The last statement  follows from the observation that at high energies
many of produced particles are concentrated around the central rapidity
\cite{prod}.
New particles with central rapidity cannot be produced as a result of only
soft parton-parton collisions. After a large momentum loss, an active
parton can hardly participate in the production on backward  target nucleons.
In the terminology of
Ref.\cite{Bodw}, the absorption of active partons can be described by hard
active-spectator interactions or collinear active-spectator interactions
with large momentum transfer $l_{+}\sim P$. We emphasize that
the proofs of the strong factorization performed in Refs.\cite{CSS}-\cite{BBL}
do not concern these two types of active-spectator interactions and the
target-length
condition  (2) was not proved for these initial state interactions.
For example, the strong factorization has not been proved when $l_{+}
\sim P/ML$ (see Eq.(2.24) from Ref.\cite{Bodw}).
The cancellation of two-step diagrams, demonstrated in Ref.\cite{Muel},
is not effective for these interactions because the
coherency of parton-parton interactions in nuclei is lost when $l_{+}$ is
much larger than the inverse internucleon distance\cite{BBL,Akul1}.
Furthermore, the limitation (3), obtained for the induced
gluon radiation, was not proved for some other initial state interactions,
for example for the elastic active-spectator scattering with large $l_{+}$.
However, it is reasonable to expect that the  contribution
of large momentum transfer collisions is small compared to total hadron-hadron
cross sections.  This limitation will be fulfilled in our model:
 the parton-nucleon absorption we introduce
manifests itself in an average  proton-proton cross section
$\sigma^{pp}_{abs}\le 1 mb$.
The large effect of the parton absorption in nuclear hadroproduction reactions
is explained by the assumed $x_{1}$-dependence of the absorption: it is
concentrated at large $x_{1}$. In some cases the
hadroproduction is sensitive to large $x_{1}$, but the contribution
of the large-$x_{1}$ region to the total hadron-hadron cross section is
small because structure functions vanish at large $x$.
Note that the parton absorption we discuss has a negligible effect for
one-nucleon targets and the strong factorization is fulfilled
in this case, in agreement with  well known experimental results.
On the other side, the violation of strong
factorization in some hadron-nucleus reactions is an experimental fact, as
it will be shown below. We will introduce a physically
motivated parametrization to fit this violation in one reaction with a better
statistics and more sensitive to the initial state absorption ($J/\psi-$
production by 800 GeV protons) in order to study nuclear effects
in all other reactions.  In other words, we will assume
that not the strong but the weak factorization is valid for nuclear targets.

{\bf II. THE MODEL}

As we argued above, the initial state parton-nucleon absorption can take
place due to
hard and semi-hard parton-parton interactions. Therefore it is reasonable
to  assume that the inclusive parton-nucleon absorption cross section is
a function of $x_{1}$, i.e. $\sigma_{abs}^{i}\equiv\sigma_{abs}^{i}(x_{1})$,
where $i=g,q$ denotes gluons or quarks.
The weak factorization of hadron-nucleus cross sections
can now be reformulated as (see Eq.(1))
\begin{equation}
f_{1}^{DY}(x_{1}) = F^{DY}(x_{1},A)f_{1} (x_{1}),
\end{equation}
where $f_{1}(x_{1})$ is the "intrinsic" projectile parton distribution and
$F$ is the factor describing the parton absorption effects.
In quarkonium production, the factor $F^{Q}$ can be also $x_{F}-$dependent
due to final state interactions of produced partons.
The final state interactions can account for possible comover
interactions\cite{Brod2} and$/$or interactions of non-singlet configurations
\cite{Weise1}. However, as we mentioned above, these final state interactions
are expected to change slowly in the positive-$x_{F}$ region we are concerned
with. Therefore we will assume that an $x-$dependence of suppression factors
$F$ is accounted for by the
$x_{1}-$dependence of $\sigma_{abs}^{i}(x_{1})$.
The final state effects in quarkonium production will be represented
by a constant cross section $\sigma_{f}$.
In our earlier work\cite{Akul1}, we have used constant cross sections
$\sigma_{abs}^{i}$ to describe initial state effects
and the results of the present paper will significantly differ from the
results of Ref.\cite{Akul1}.

In the case of the square well form of nuclear density, the factor $F^{Q}$ for
the quarkonium production ($\sigma_{f}\neq 0$) is given by the eikonal form
\begin{equation}
F^{Q}_{i}(x_{1})=\frac{3}{\sigma_{f}-\sigma_{abs}^{i}(x_{i})}\lbrace
\frac{\sigma_{abs}
^{i3}(x_{1})}{c_{abs}}[1-e^{-c^{i3}_{abs}}(1+c_{abs}^{i})] -
\frac{\sigma_{f}}
{c^{3}_{f}}[1-e^{-c_{f}}(1+c_{f})]\rbrace,
\end{equation}
where $c_{abs,f}\equiv 2\rho \sigma_{abs,f}R_{A}$ and the nuclear density
and radius are denoted by $\rho$ and $R_{A}$.
For the DY production ($\sigma_{f}=0$), this
equation can be rewritten as
\begin{equation}
F^{DY}_{i}(x_{1})=\frac{3}{c^{i3}_{abs}}[\frac{c^{i2}_{abs}}{2} +
e^{-c_{abs}^{i}}(1+
c_{abs}^{i})-1].
\end{equation}
In Eqs.(6) and (7) we assumed that the exclusive production cross section is
much smaller than $\sigma_{abs}^{i}$ or $\sigma_{f}$.

We have to determine the functions $\sigma_{abs}^{i}(x_{1})$.
The physical hadron-hadron absorption cross section is given by
\begin{equation}
\sigma^{hh}_{abs} = \sum_{i=g,u,d\ldots}\int dx_{1} \sigma_{abs}^{i}(x_{1})
f^{i}(x_1).
\end{equation}
Since parton distributions diverge as $x\rightarrow 0$, we must assume
$\sigma_{abs}^{i}(x_{1}=0)=0$ in order to obtain a finite $\sigma_{abs}^{hh}$.
A phenomenological analysis of the $J/\psi-$suppression has shown that the
data are better discribed when $d\sigma_{abs}^{i}(x_{1}=0)/dx_{1}=0$ as well.
In this case we must assume that $d\sigma_{abs}^{i}(x_{1}=1)/dx_{1}=0$.
This condition can be obtained from Eq.(8) using the probabilistic
interpretation of parton distributions $f^{i}(x)$.
Taking into account these limitations, we
found that the functions
\begin{equation}
\sigma^{i}_{abs}(x_{1})=\sigma_{max}^{i} sin^{6}(x_{1}\frac{\pi}{2}),
\end{equation}
fit well the empirical $x_{F}-$dependence of the  $J/\psi-$
suppression in proton-nucleus collisions at 800 GeV.
We will use the form (9) to derscribe all reactions.
The function $sin^{6}(x\frac{\pi}{2})$ is shown in Fig.1. The parameters
$\sigma_{max}^{i}$ characterize the strength of absorption and
are  extreme values of hadron-hadron absorption cross sections
when the projectile hadron is composed of only one
parton $i$ (when $x_{1}=1$ for this parton).
It is clear from Eq.(8) and Fig.1 that the physical value
$\sigma_{abs}^{hh}$
is much smaller than $\sigma_{max}^{i}$ ($\sigma_{abs}^{hh}/\sigma_{max}^{i}
\sim 2\%$). We may assume $\sigma_{max}^{q} = \frac{4}{9}\sigma_{max}^{g}$
for all light quark flavors because of the standard color factors.
We now have to fix the projectile dependence of $\sigma_{max}^{i}$.
At high energies, a pion-hadron cross section is about one half of the
corresponding proton-hadron
cross section. Since $\sigma_{max}^{i}$ have the meaning of extreme values
of hadron-hadron cross sections, we extend the above relation between
pionic and nucleonic cross sections on $\sigma_{max}^{i}$ as well.
For definiteness, the projectile dependence will be represented by
$\sigma_{max}^{i}(pions) = 0.4 \sigma_{max}^{i}(protons)$.
As a result, we have only one free parameter, say $\sigma_{max}^{g}(protons)$,
to describe the initial state effects.
We will use
\begin{equation}
\sigma_{max}^{g} (protons) = 30 mb.
\end{equation}
The remaining parameters ($\sigma^{q}_{max}(protons)$ and $\sigma^{q,g}_{max}
(pions)$) can now be calculated. The parametrization (9) and (10) will
{\it not} be
adjusted for different projectile energies or for different reactions.

For the quarkonium production, we have to take into account the final
state interactions. In the study of the empirical regularities in
the nuclear charmonium suppression
\cite{Hufner}, it was found that the $x-$independent part of nuclear
suppression can be accounted for by $\sigma_{f}\approx$ 5.8 mb.
In our model, these value will simulate the
$x-$independent final state interaction of produced heavy quarks.
A self-consistent description of
charmonium and dilepton production will prove {\it a posteriori} that
in our model there is no room for a significant $x_{F}-$ dependence
of $\sigma_{f}$ in the region $x_{F}\ge 0.2$.
Final state effects have indeed a very modest $x_{F}-$dependence in
some models (see, e.g., Ref.\cite{recomb2}). The fact that a quite large
value of $\sigma_{f}$ fits the final state interactions may indicate
that heavy quark-antiquark pairs pass through the nucleus as non-singlet
configurations, in agreement with the conjecture of Ref.\cite{Weise1}.
In fact, $\sigma_{f}$ is almost three times larger than the geometrical-size
estimation for the $J/\psi-$proton total cross section\cite{size}.
A similar result for the final state $c\overline{c}-$nucleon cross section
has been obtained in some other models\cite{recomb2}.
We think that the symmetrical
description of initial and final state interactions in quarkonium production,
adopted in Ref.\cite{Eloss},  is not justified. There can be an important
difference
between the projectile hadron and the produced heavy quark configuration,
which is not yet hadronized during its propagation in nuclear matter. As we
will
show, the price for the charm suppression fit of Ref.\cite{Eloss} is the
underestimation by that model of initial state effects in the DY production
by pions.

{\bf III. CHARMONIUM PRODUCTION}

To calculate the charmonium production cross section, we take into account
both the gluon-gluon fusion and quark-antiquark annihilation subprocesses.
{}From the numerical results, we concluded that both channels are important
in the considered $x_{F}-$region and the relative contribution of these
two channels is in agreement with the result of Ref.\cite{recomb2}.
The proton-nucleus cross section is given by
\begin{equation}
\frac{d\sigma}{dQ^{2}dx_{F}} =
F^{Q}_{g}\frac{\sigma^{gg}x_{1}x_{2}}{Q^{2}(x_{1}
+x_{2})}g_{1}(x_{1})g_{2}(x_{2})+F^{Q}_{q}\frac{\sigma^{qq}x_{1}x_{2}}
{Q^{2}(x_{1}+x_{2})}\sum_{f=u,d}[q_{1}^{f}(x_{1})\overline{q}_{2}^{f}(x_{2})
+\overline{q}^{f}_{1}(x_{1})q^{f}_{2}(x_{2})].
\end{equation}
The hard partonic cross sections $\sigma^{ii}$ are taken from Ref.\cite{part}.
It is known that
the nuclear structure functions, $g_{2}(x_{2}), q_{2}(x_{2})$
and $\overline{q}_{2}^{f}(x_{2})$, are not equal to free nucleon structure
functions. Nuclear structure functions measured
in deep inelastic scattering have two pronounced pecularities\cite{NMC}:
the shadowing at very small $x_{2}$
and the depletion (the EMC-effect) at $x_{2}\ge$0.4
(the antishadowing at $x_{2}\sim 0.1$ is very small and will not be taken
into account).
The similar effects
may be expected in the nuclear gluon distributions.
In this section, we will not take into account these nuclear effects for the
following reasons.
If the nuclear shadowing at small $x_{2}$ is due to projectile initial
state interactions in nuclear matter (in the case of electroproduction
at small $x_{2}$ the projectile can be a virtual quark or meson\cite{shad}),
then these effects are
already included in factors $F$. We will argue that the alternative
interpretation of nuclear shadowing, assuming the parton recombination
in nuclei, is inconsistent with the charmonium production data.
The EMC-effect is important  at  large $x_{2}$,  far from the region
$x_{2}\le 0.14$ probed
by the current experiments\cite{Badier,Alde1}.
The latter argument is not correct for the quarkonium production in the
region $x_{2}\sim 0.3$ studied in some recent experiments\cite{Alde3}.
At such $x_{2}$ the EMC-effect in nuclear gluon and quark distributions
can be noticeable. In this section we use the free proton structure functions
from Ref.\cite{Eich} for both $f_{2}(x)$ and $f_{1}(x)$.

The $x_{2}-$ and $x_{F}-$dependence of nuclear effectiveness $\alpha$
($\alpha\equiv ln(\sigma_{A}/\sigma_{p})
/ln(A)$) for 800  and  200 GeV protons is shown in Figs.2 and 3.
The agreement with the data at 800 GeV (the solid lines and the diamonds)
is not surprising because we used this set
of data to find the parametrization (9) and (10).
The same for 200 GeV protons is shown  by the short-dashed lines and crosses.
This result is already non-trivial, espessially the
$x_{2}-$dependence of $\alpha$ at 200 GeV.
The net effect of final state interactions is represented by
the solid and short-dashed lines in the small $x_{F}-$region in Fig.3.

The parton-recombination model prediction is shown by the dotted line in
Fig.2.
In this case, we have calculated  $\alpha$ using the quark and gluon
nuclear distributions from Ref.\cite{recomb2} (the set 1  for gluons)
and the constant value $\sigma_{f}=5.8 mb$ to approximately account for the
final state effects. Note that the dotted line is expected to describe
the data for both energies, within a kinematically allowed region of
$x_{2}$ for each energy, because the difference between the results for
two energies is less than 1\% at the same $x_{2}$.
We can see from Fig.2 that the
parton-recombination model is inconsistent with the data at both energies.
Moreover, the parton-recombination model is ruled out
by the data, if the origin of the nuclear suppression
and its gross $x_{F}-$dependence are the same at
both proton energies (this is very likely given the results
presented in Fig.3).  In fact, in the whole $x_{2}-$region measured
at 200 GeV, $0.037<x_{2}<0.14$,  the  effect of the parton recombination
is negligible. The same should be valid at 800 GeV in the region
$0.01<x_{2}<0.04$ since these two $x_{2}-$regions correspond to the
same $x_{F}-$region, taking into account the proton energy.
. It follows from Fig.2 that the dotted line is
in conflict with this expectation.
Therefore the parton recombination model not only underestimates the
charmonium suppression but also demonstrates a wrong $x_{2}-$dependence
of the suppression at different projectile energies.
This can be corrected only if the parton recombination model
will be combined with some other model of nuclear suppression, which
predicts a specific dependence on the projectile energy at fixed
$x_{F}$. For example, the EL-model does not have this property and cannot
restore the correct $x_{2}-$dependence, if being combined with the parton
recombination model. We may assume that the nuclear parton recombination
is less important than it is usually believed and will not take
into account this effect in what follows.

Though our solid and short-dashed lines in
Figs. 2 and 3 correctly reproduce the trend of the data,
our results underestimate $\alpha$ at $x_{2}>0.06$.
This discrepancy is eliminated if we assume that there is a
20\%-enhancement of gluons in tungsten at $0.06<x_{2}<0.15$.
The calculated $\alpha$ with excess gluons included is shown by the
long-dashed lines in Figs.2 and
3. The resulting $\chi^{2}$ of our calculations is about 1 at both proton
energies.
Note that this excess gluon contribution affects $\alpha$ only at 200 GeV.
The gluon enhancement we assumed is in qualitative agreement with the
electroproduction data\cite{Amaud}. The momentum fraction carried by
the excess gluons is about 3\% for tungsten and is in qualitative agreement
with the earlier estimations\cite{SF}. It may be assumed
that the excess gluons play the role the excess pions have been expected
to play - they are responsible for the nuclear binding\cite{SF}.
Observe that the distances involved by the  excess gluons are of the
order of internucleon distances in nuclei. It is also remarkable that the
momentum fraction carried by excess gluons is comparable with the parton
momentum fraction, lost because of the nuclear binding\cite{AKV}.
Is it an
indication on the gluonic origin of nuclear forces?  We will come back to
this problem in Sec.V. Note that the excess gluons are not the only
possible explanation of the above discrepancy. For example, a similar
effect can be reproduced by  an energy dependence of final state interactions.

The projectile dependence of the charmonium suppression is illustrated
in Fig.4.
The $x_{F}-$dependence of $\alpha$ for 200 GeV $\pi^{-}-$mesons
without excess gluons (long-dashed line) is compared to the data
for pions at this energy (diamonds).
The same for protons  at 200 GeV
is shown by the short-dashed line and by crosses.
It follows from
this figure that our model correctly reproduces the projectile dependence
of nuclear suppression.
Remember that in our model, the projectile dependence is represented
by the ratio $\sigma_{max}^{i}(pions)/\sigma^{i}_{abs}(protons)$, introduced
in Sec.II.
However, the long-dashed line still overestimates
the suppression. The agreement with the data is improved when
the excess gluon contribution is included (the solid line) in the same way
as for proton-nucleus collisions.

To summarize, the absorption of initial partons
can provide a self-consistent description of $x_{2}-$ and $x_{F}-$dependences
of nuclear charm suppression at different energies and for different
projectiles. In the region $0.06<x_{2}<0.15$, the agreement with the
data is improved by  the inclusion of excess gluons.

{\bf IV. DY PRODUCTION BY PIONS}

In the dilepton production by pions, the dominant contribution is due to
quark-antiquark
annihilation subprocess. Since projectile pions contain valence quarks and
antiquarks, this reaction probes essentially the same target structure
functions as the deep inelastic lepton scattering.
In the case of exact strong
factorization, the ratios $R$ of nuclear to nucleon cross sections should
coincide for these two reactions.
Therefore the DY production by pions is an excelent test of strong
factorization
for nuclear targets. We now show that the weak factorization
is more consistent with the data than the strong factorization.
The cross section for producing dilepton pairs in collisions of pions
and nuclei is given by
\begin{equation}
\frac{d\sigma^{DY}}{dQ^{2}dx_{F}} = F^{DY}\frac{\sigma^{qq}x_{1}x_{2}}
{Q^{2}(x_{1} +x_{2})}\sum_{f=u,d,s}[q_{1}^{f}(x_{1})\overline{q}_{2}^{f}(x_{2})
+ \overline{q}_{1}^{f}(x_{1})q_{2}^{f}(x_{2})].
\end{equation}
The suppression factor $F^{DY}$, given by Eq.(7), was calculated as described
in Sec.II.
The data for this reaction\cite{Bordalo} covers a large region of $x_{1}$
and $x_{2}$ for two pion energies, 140 and 286 GeV. Therefore
we now have to take into account the nuclear modification of structure
functions $q_{2}^{f}(x_{2})$ and $\overline{q}^{f}_{2}(x_{2})$.
This modification will be represented by the depletion of nuclear structure
functions at medium $x_{2}$ (the EMC-effect). The form of this depletion
that we
used in numerical calculations reproduces the electroproduction data for
nuclear targets\cite{NMC}.
As it was discussed above, in the present model there is no physical reason
to take into account the shadowing
of nuclear structure functions at very small $x_{2}$ since  such effects
are assumed to be included in $F^{DY}$.
We used the free nucleon structure functions from Ref.\cite{Eich}.
In numerical calculations, we integrated over the  $Q^{2}-$ and $x_{F}-$
regions, measured in the current experiment\cite{Bordalo}.

The $x_{2}-$dependence of the ratio $R$ for the tungsten
at 140  and  286 GeV  is shown
in Figs. 5 and 6, respectively.
Our results are represented by the solid lines, the predictions of the
EL-model by the long-dashed lines and the strong factorization
result (the EMC-ratio) by the short-dashed lines.
Here and hereafter the EL-model predictions are calculated using Eq.(4)
with $\kappa=0.003$ and $n=2$.
The strong factorization is completely inconsistent with the data at
140 GeV ($\chi^{2}=5$).
The EL-model ($\chi^{2}=2.3$ at 140 GeV) could explain the data
with a larger parameter $\kappa$, but
this would lead to a futher violation of the limitation (3).
Our results are in a reasonable accord with the data at 140 GeV ($\chi^{2}=
1.1$).

The Fig.6 is less conclusive because the error bars in the most
important region $x_{2}\ge 0.3$ are larger than the difference between the
theoretical predictions and $\chi^{2}<2$ for all models.
By comparing Figs. 5 and 6,
we conclude that in this reaction it is  confusing to combine the data points
for different energies because the
$x_{2}-$dependence of $R$ for two energies is very different.
For example in the region $0.1<x_{2}<0.3$, the suppression at 140 GeV
is much stronger than at 286 GeV.
The "diffraction" minimum at $x_{2}\approx 0.37$ in our predictions
for $E_{\pi}=$ 286 GeV takes place  because the lower mesured mass region,
$(4.2 GeV)^{2}<Q^{2}<(8.5 GeV)^{2}$, is becoming kinematically forbidden
at that $x_{2}$.  Observe that our model predicts a fast depletion at
small $x_{2}$ because larger  values of $x_{1}$ contribute in this region.
New data at $x_{2}\le 0.1$ could help to test our model. As it will be
shown in the next section, such depletion at small $x_{2}$ has
been observed in the DY production by protons.

The $x_{1}-$dependence of $R$ is shown in Figs. 7 and 8.
Taking into account the very large uncertainty in determining $x_{1}$
for the data points
(not represented in these figures but shown in Ref.\cite{Bordalo}),
it is hard to make a preference between the models. A better statistics is
needed to make a definite conclusion about the $x_{1}-$ dependence of $R$.

The dependence of $R$ on the mass of the produced dilepton pair is shown
in Figs. 9 and 10. The data at 140 GeV  demonstrate  a sooth
behaviour and Fig.9 is  quite conclusive: the present model ($\chi^{2}=0.5$)
provides a much better agreement with the data than the EL-model
($\chi^{2}=2.6$)
or the strong factorization ($\chi^{2}=4.5$).
Note that the EL-model with $n$=2 could not fit the observed
mass dependence of the suppression even
with a larger $\kappa$. The point is that in the EL-model, the suppression
decreases with produced mass because of the factor $(Q_{0}/Q)^{n}$ in (4)
and this contradicts the trend of the data in Fig.9.
The data at 286 GeV have large error bars in the
most important region $(x_{1}x_{2})^{1/2}\ge 0.5$ and all theoretical
curves have $\chi^{2}<1.5 $ at this energy.
However, it is our impression that our results are in a reasonable accord with
the data at 286 GeV.

Let us summarize the results of this section.
The Figs. 6,7,8 and 10 make no definite preference between the
models.
But the Figs. 5 and 9 allow us to conclude that the present model
describes the DY production  by pions much better than the EL-model.
These two figures also present an evidence that the strong
factorization is violated in nuclei.

{\bf V. DY PRODUCTION BY PROTONS}

As in the case of the DY production by pions, the DY production by protons
is dominated by the quark-antiquark annihilation subprocess. The difference
between these two reactions is that protons probe mainly the antiquark
content of a target, because protons have no valence antiquarks.
Therefore this reaction is sensitive to a possible contribution of excess
mesons in nuclei. The cross section for producing dilepton pairs in
proton-nucleus collisions is given by Eq.(12). We have used the free nucleon
structure functions as $q_{2}^{f}(x)$  and $\overline{q}^{f}_{2}(x)$
for the following reason. The nuclear modification of the antiquark
distribution is unknown, but from the electroproduction data\cite{NMC}
it follows that the EMC-like nuclear effects should be unimportant in the
region $x_{2}<0.3$, measured in the current experiment with 800 GeV protons
\cite{Alde2}. The nuclear shadowing at small $x_{2}$ is not included
in structure functions for the same reason as in the previous sections.
Thus, in this reaction
we will study the net effect of the initial state absorption represented
by the factor $F^{DY}$. In the previous work\cite{Akul2}, we have considered
the same reaction  with the constant parton absorption cross section.
In that paper, we have concluded that the excess pion contribution with
the average excess pion number per nucleon $n_{\pi}\approx 0.07$  is
consistent with the data. We will now show that the
$x_{1}-$dependence of the absorption cross sections, introduced in the
present paper,
drastically changes the role of initial state absorption in this reaction
and affects the conclusion of Ref.\cite{Akul2}.

We used the free nucleon structure functions from Ref.\cite{Eich}.
We have integrated over the same mass region as in the experiment\cite{Alde2}.
The $x_{2}-$dependence of the ratio $R$
for the iron and tungsten targets is shown in Fig.11 by the solid and
dashed lines, respectively. An important
result is that the suppression is concentrated mainly at $x_{2}<0.05$.
The origin and the location of this depletion are exactly the same as in the
case
of the charmonium  production by protons at 800 GeV (the solid line in Fig.2).
This observation is supported by the fact that the calculated $R$ from Fig.11
is in accord with the data
($\chi^{2}=0.7$). The A-dependence of the suppression at low
$x_{2}$ is also well reproduced by our model. In this situation, there is no
room for any noticeable contribution of excess pions, at least at
$x_{2}\ge 0.04$. If our present model is correct, than the upper
limit for the momentum fraction carried by excess pions is about 1\%.
This conclusion is qualitatively consistent with the results of
Ref.\cite{Shakin} that the structure functions of off-mass-shell pions
are significantly depleted.
In this case the excess gluons, considered in Sec.III, are another
possible candidate to carry the missing parton momentum in nuclei\cite{AKV}.

The $x_{F}-$dependence  of $R$ for the iron is shown in Fig.12.
Though the qualitative trend of the data is more or less reproduced
by the theoretical curve, the suppression at $x_{F}>0.5$ is overestimated
by our model. In that region, the values $x_{2}\le 0.04$ have the
noticeable contribution. We may assume  that in the region $x_{2}\le 0.04$,
not covered by the data points in Fig.11, either the nuclear structure
functions are enhanced due to some excess particles or our model overestimates
the suppression. In any case, our model leaves no room for
the parton recombination effects in this reaction, as well as in the
charmonium production.

{\bf VI. DISCUSSION}

We have assumed a new type of initial state effects, the $x_{1}-$dependent
absorption of fast initial partons. As we argued in Sec.I  and
in Ref.\cite{Akul1}, such effects in nuclei are not {\it a priori} excluded
by the factorization theorem forecasts.  Using only one physically
motivated empirical function for the absorption cross sections,
we have qualitatively explained nuclear effects
in several reactions  with fast protons and pions. We claim that
in the currently discussed experiments on charmonium production,
the $x-$dependence
of nuclear suppression can be accounted for by the initial state interaction
we assumed. In this case, the final state effects in the region $x_{F}>0$
can be fitted by the constant cross section  $\sigma_{f}$.
With the excess gluon contribution included, the present model can explain
also the energy and projectile dependence of the charmonium suppression.
The dilepton production suppression is also
qualitatively explained, though our model sometimes overestimates the
suppression. This possible discrepancy can be removed by  adjusting
the parameters, for example by reducing $\sigma_{max}^{q}/\sigma_{max}^{g}$.
The present results
are obtained using a minimum of arbitrary assumptions and allow us to conclude
that this model is able to self-consistently describe nuclear effects
in different hadroproduction reactions.  The
$\Upsilon-$ and $J/\psi-$production at low and negative $x_{F}-$ demands the
incorporation of comover interaction and nuclear modification of gluon
distribution and will be considered somewhere else.

If the present model is correct, then the following
qualitative conclusions can be made. 1) The strong factorization is violated
by the initial state interactions  in hadroproduction on nuclei. The
weak factorization takes place instead. 2) The parton recombination
has no significant contribution to charmonium suppression.
3) There is an indication of the presence of excess gluons
in nuclei at $0.06<x_{2}<0.15$, carrying about 3\% of the total momentum.
4) The energy-loss model cannot self-consistently describe the quarkonium
and dilepton production if the mass dependence of the energy loss
is given by $1/Q^{2}$. 5) There is no indication of the excess pion
contribution in dilepton production in proton-nucleus collisions.
The fraction of the nuclear momentum, lost by partons due to
nuclear binding,  can be carried by the excess gluons. This point
illustrates the difference between the binding correction\cite{AKV}
and pionic models\cite{Magda} for the EMC-effect:
the former is not based on the presence of excess pions in nuclei.

The above conclusions do not mean that the soft initial state interactions,
assumed by the EL-model, take no place in nuclei. Instead, these interactions
may be responsible for the observed $p_{T}-$dependence of hadroproduction
on nuclei. But the role of these interactions in the $x-$dependence
of nuclear suppression can be minor, in agreement with the limitation
(3). The validity of the present model can be tested, e.g., by remeasuring
with a better accuracy the DY production by pions, where we predict
several specific pecularities.
We believe that the results of this paper can be useful for an analysis
of nuclear effects in hadro- and electroproduction, even if our present
model will be ruled out by future studies.

{\bf ACKNOWLEDGMENTS}

I would like to thank the Research Institute for Particle and Nuclear
Physics, Budapest, for  the warm hospitality.

This work was funded by the Russian Foundation of Fundamental
Research (contract No 93-02-14381).

\newpage

{\bf Figure captions.}

FIG.1.    The function $sin^{6}(x\pi/2)$, representing the $x-$dependence
of parton-nucleon absorption cross sections.

FIG.2.    The nuclear effectiviness $\alpha$ versus $x_{2}$ for
charmonium production in proton-nucleus collisions at the
energies 800 GeV (solid line) and 200 GeV (short-dashed line). The latter
with excess gluons included is represented by the
long-dashed line.
The parton-recombination model prediction is shown by the dotted line.
The data at 800 GeV (diamonds) and 200 GeV (crosses)
are from Refs. \cite{Alde1} and \cite{Badier}, respectively.

FIG.3. The same as in Fig.2 versus $x_{F}$  (without the parton-recombination
model prediction).

FIG.4. $\alpha$ versus $x_{F}$ for the charmonium
production in negative pion-nucleus collisions at 200 GeV (long-dashed line).
The same with excess gluons is shown by the solid line. $\alpha$
for 200 GeV protons without excess gluons is shown by the short-dashed line.
The data (diamonds for pions and crosses for protons)
at 200 GeV are from Ref.\cite{Badier}

FIG.5. Ratio of the dilepton yield for tungsten to proton versus
$x_{2}$ in pion-nucleus collisions at 140 GeV: the solid line is the present
model result, the long-dashed line is the result of the energy-loss model
\cite{Eloss} and the short-dashed line is the strong factorization result.
The data are from Ref.\cite{Bordalo}.

FIG.6. The same as in Fig.5 at 286 GeV.

FIG.7.  The same as in Fig.5 versus $x_{1}$.

FIG.8.  The same as in Fig.5 versus $x_{1}$ at 286 GeV.

FIG.9. The same as in Fig.5 as a function of dimensionless dilepton mass.

FIG.10. The same as in Fig.5 as a function of dimensionless dilepton mass
at 286 GeV.

FIG.11. Ratio of the dilepton yields for protons
at 800 GeV for iron (solid line) and tungsten (dashed line) as a function of
$x_{2}$.
The data for iron (diamonds) and tungsten (crosses) are from Ref.\cite{Alde2}.

FIG.12. Ratio of the dilepton yields for protons  at 800 GeV for iron
versus $x_{F}$.
The data are from Ref.\cite{Alde2}.

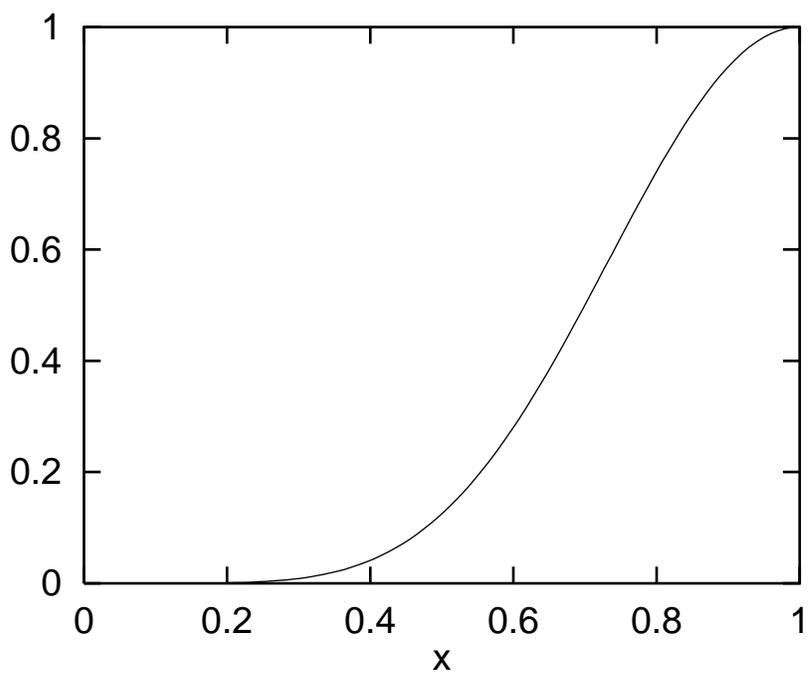
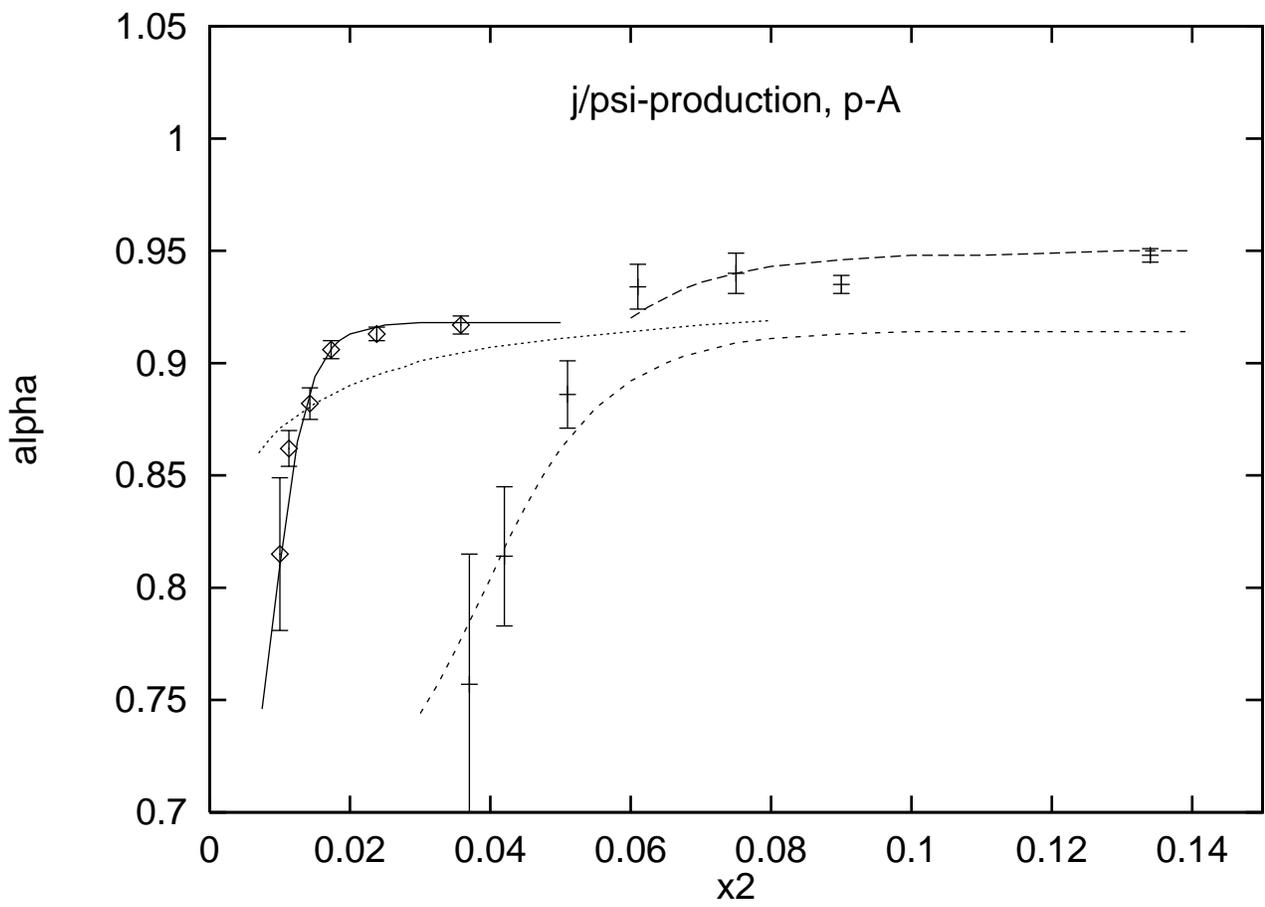
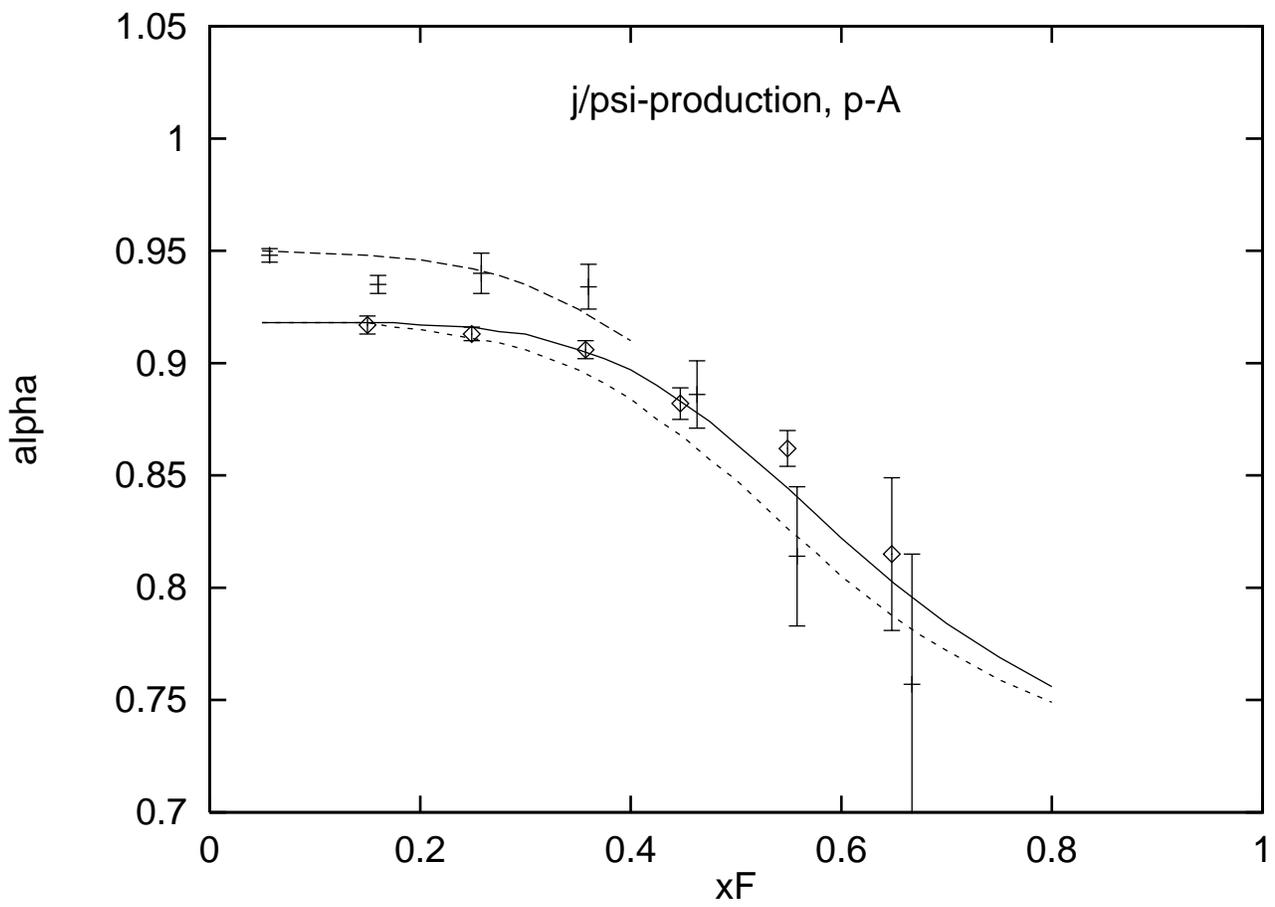
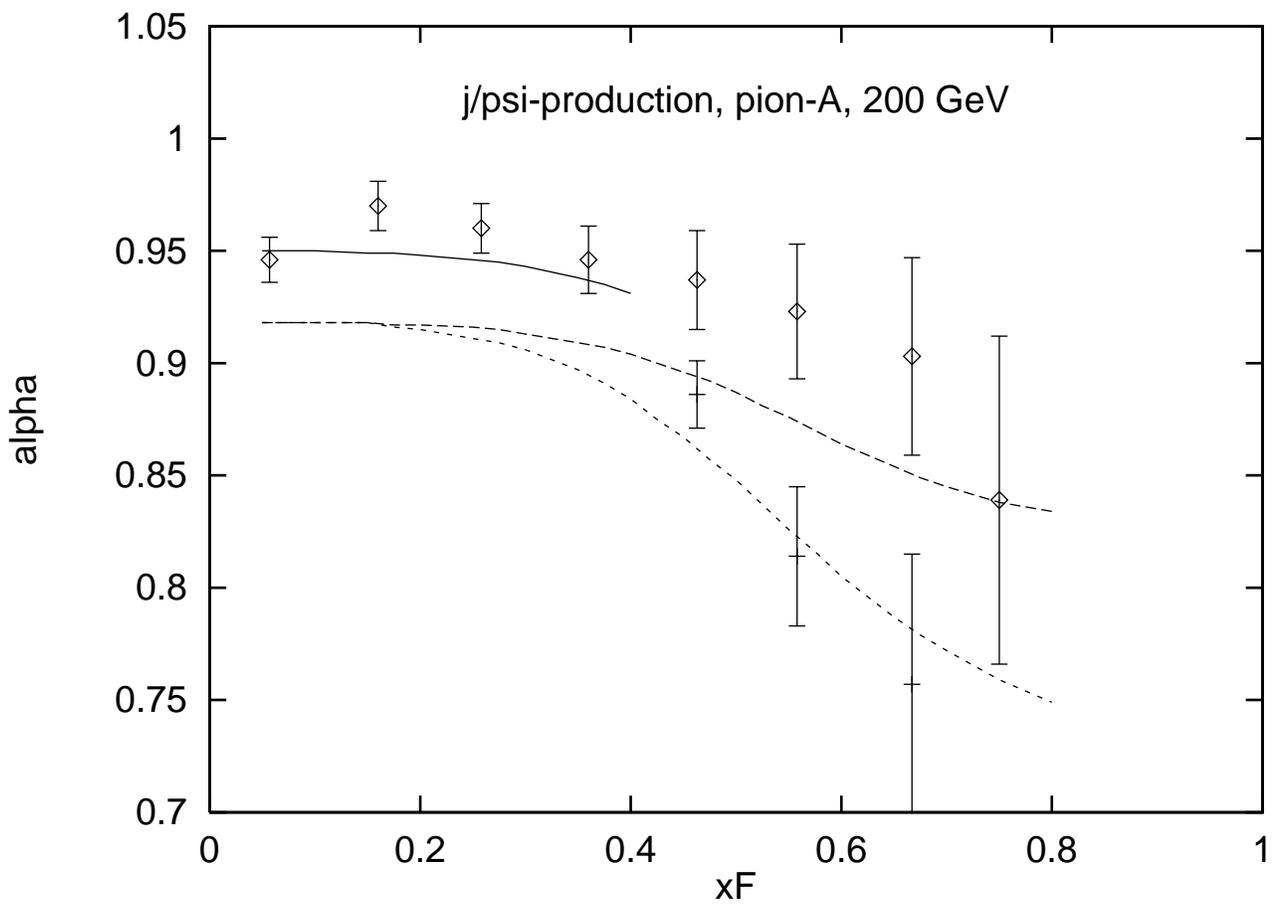
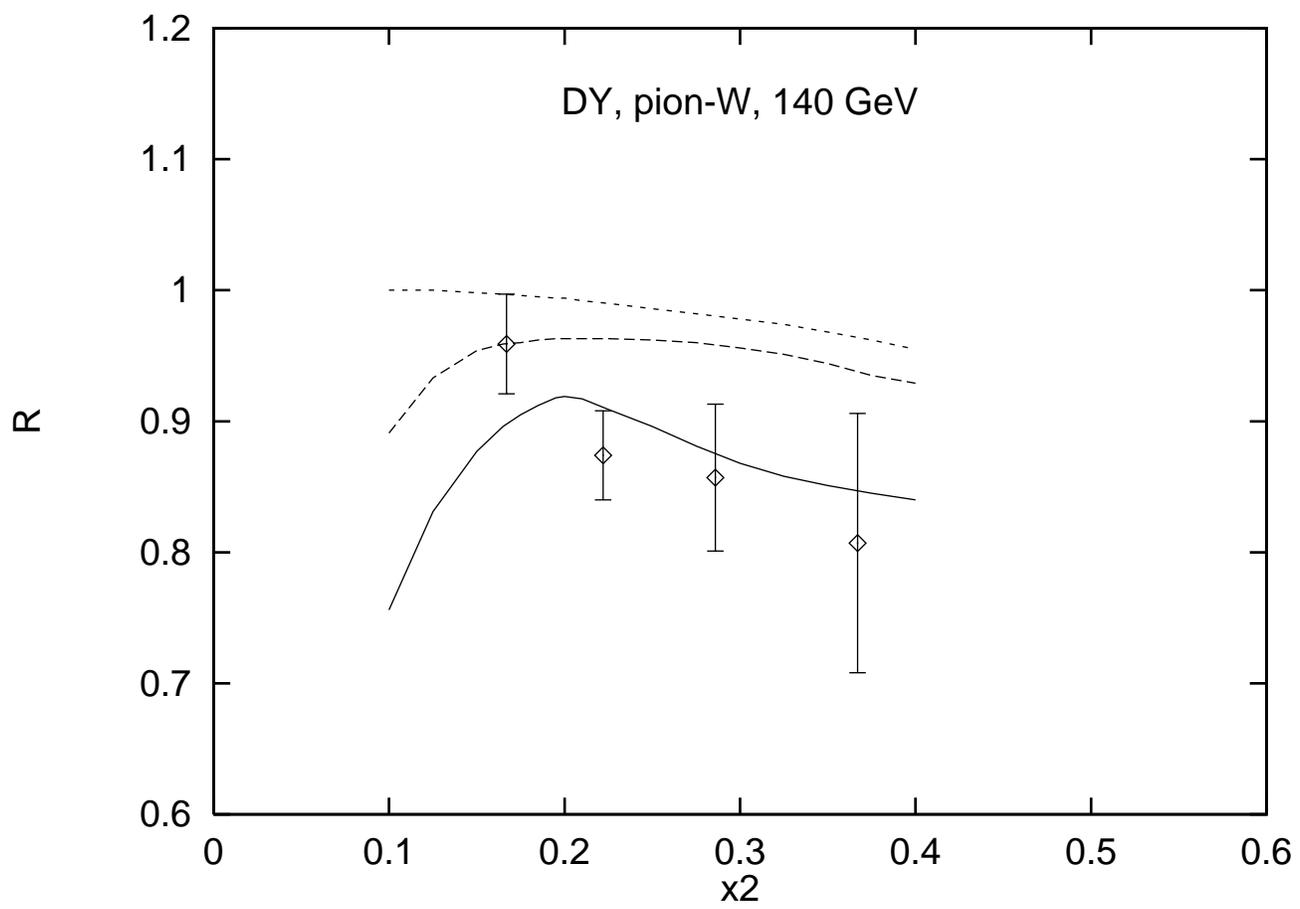
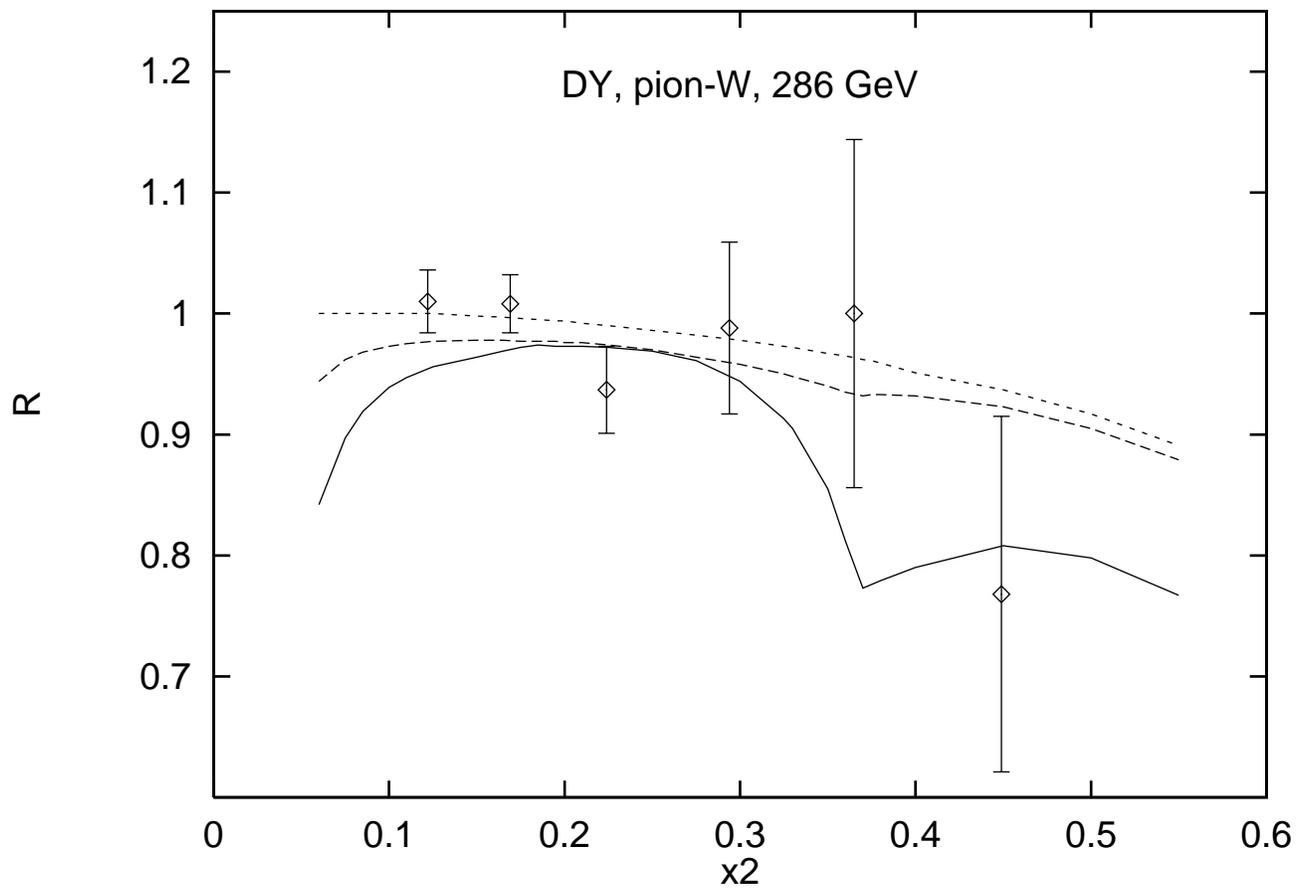
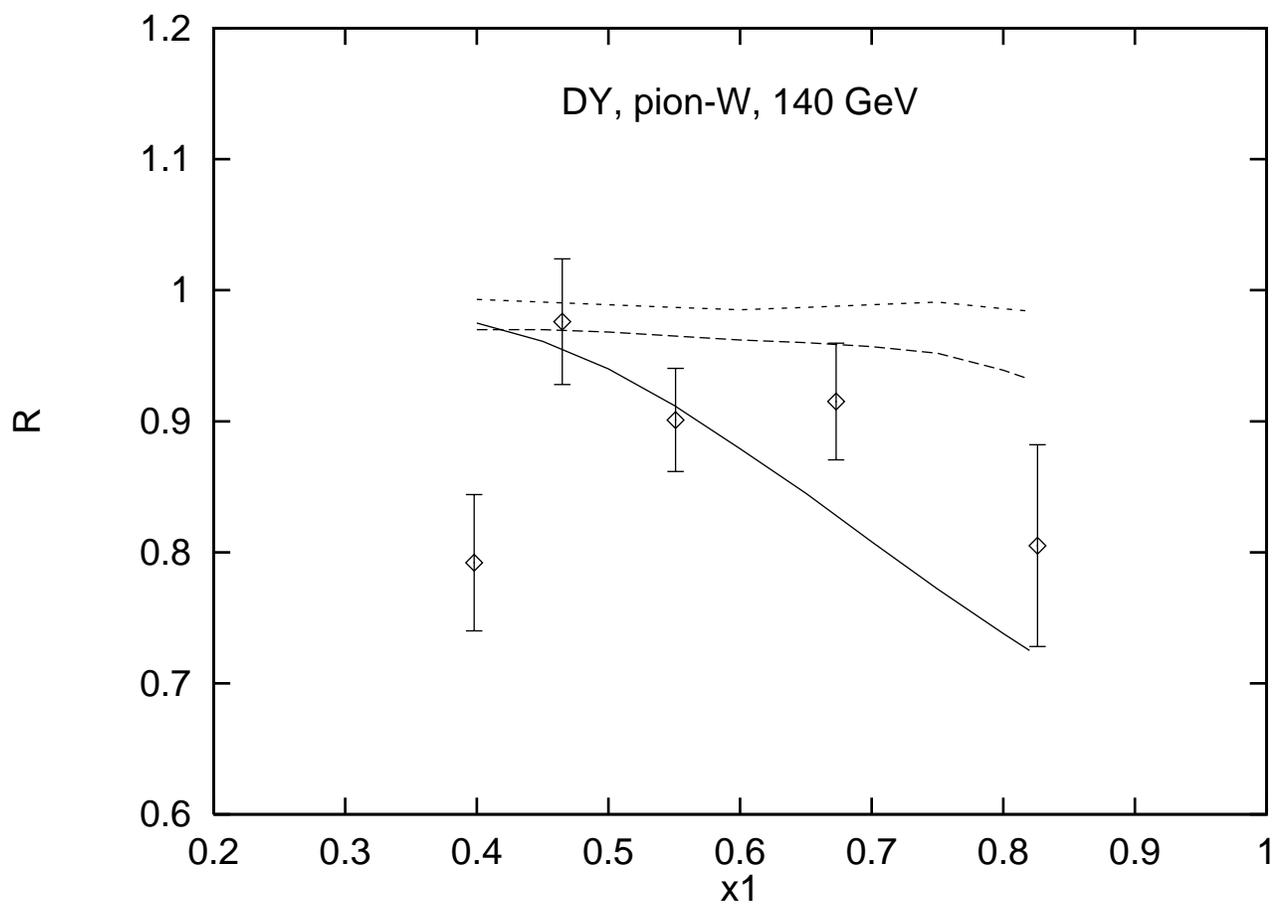
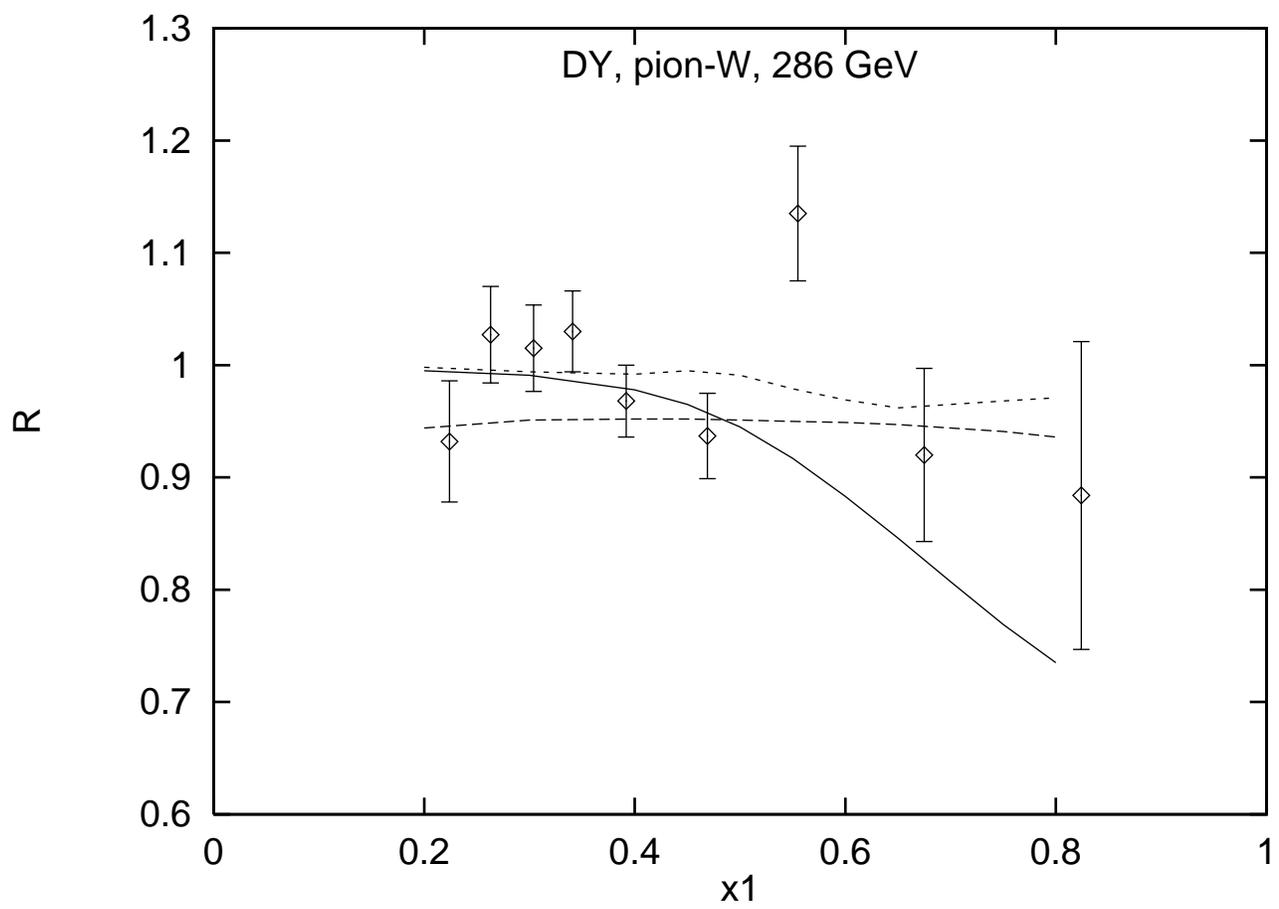
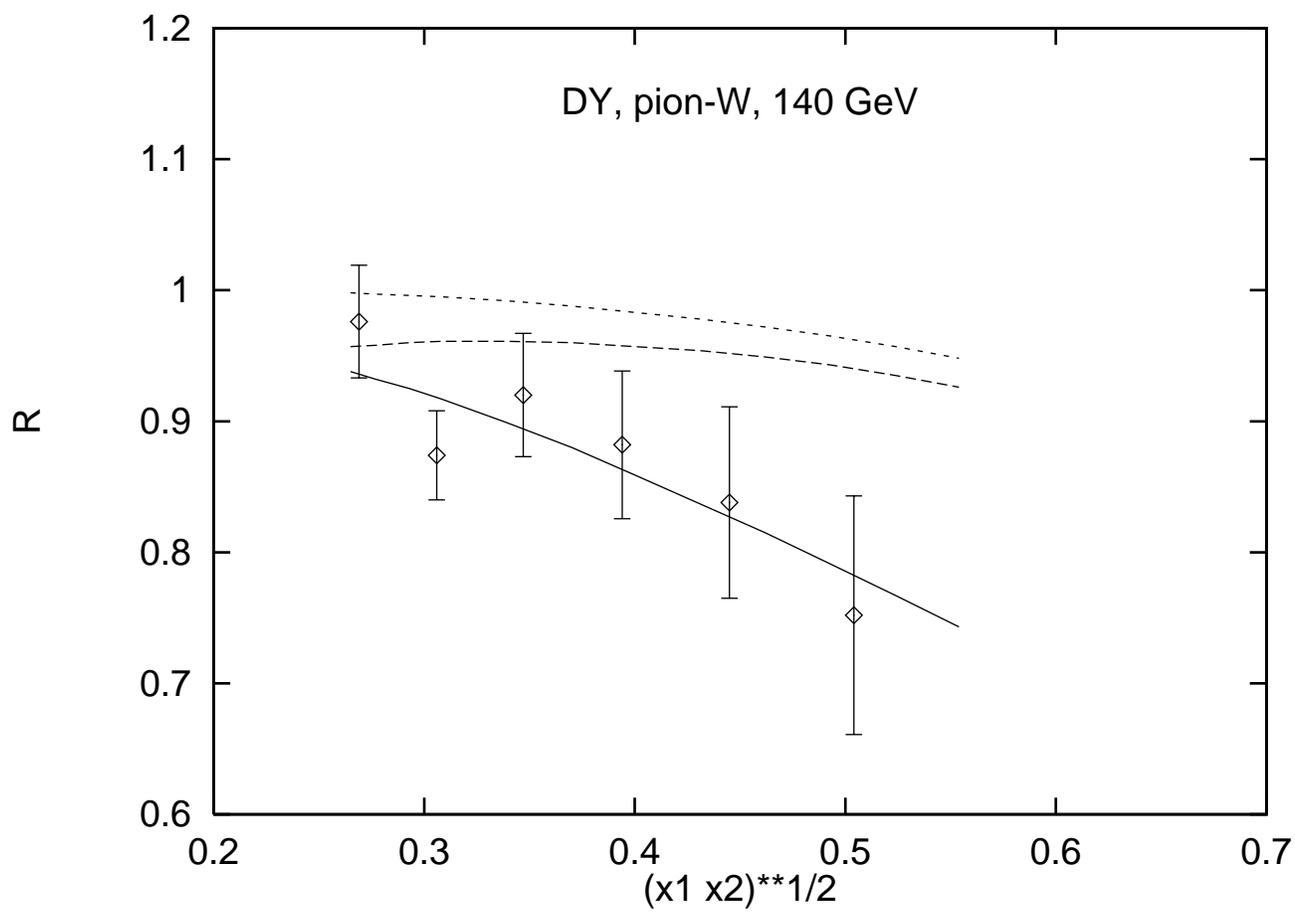
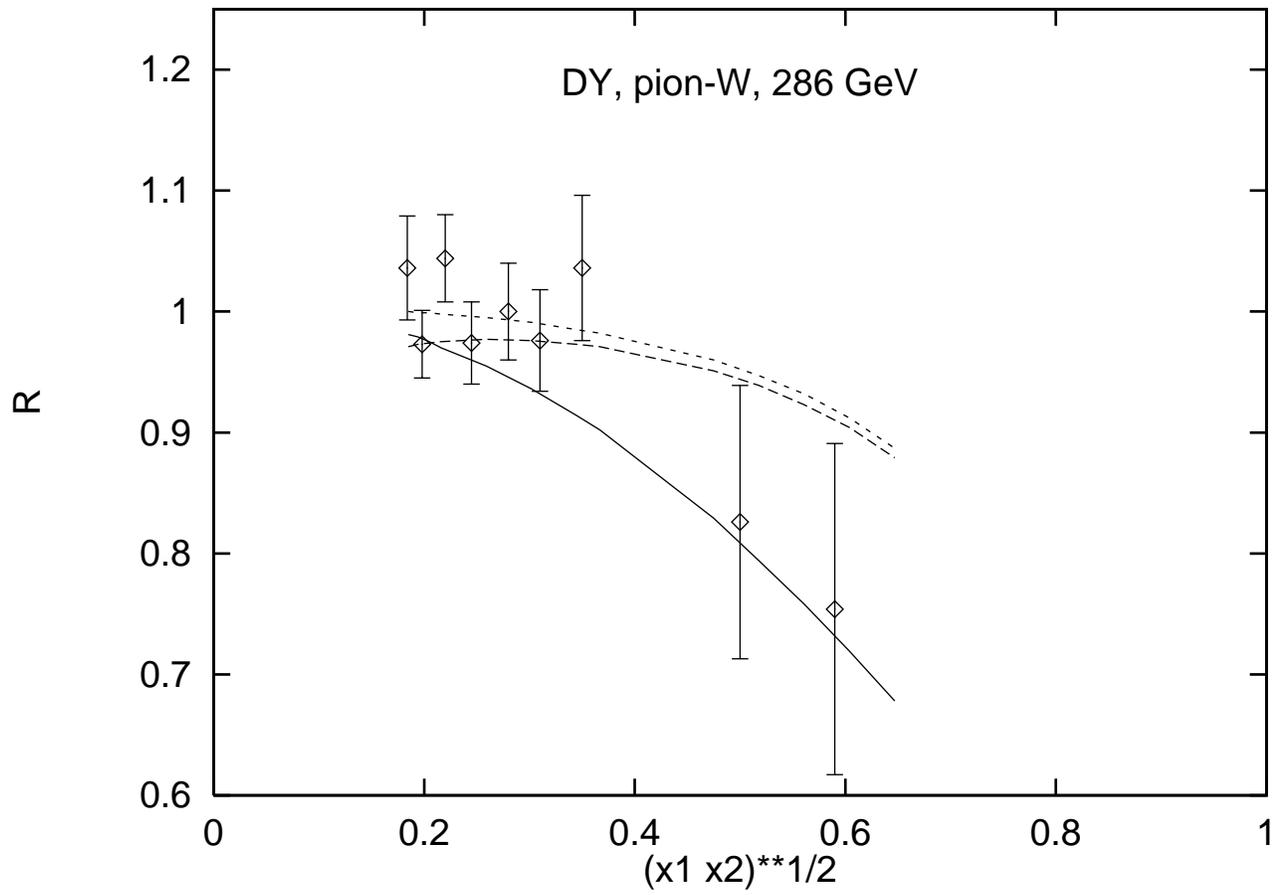
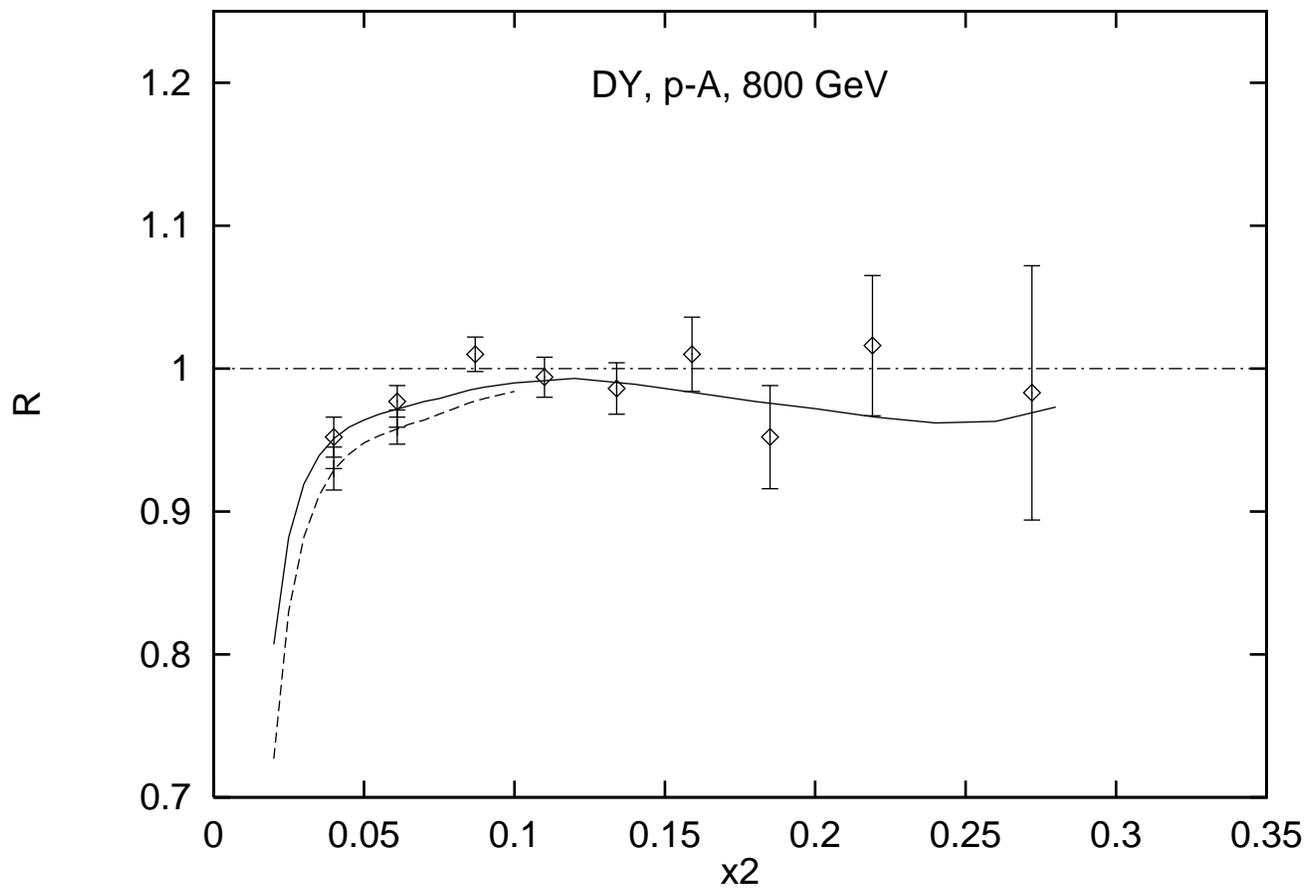
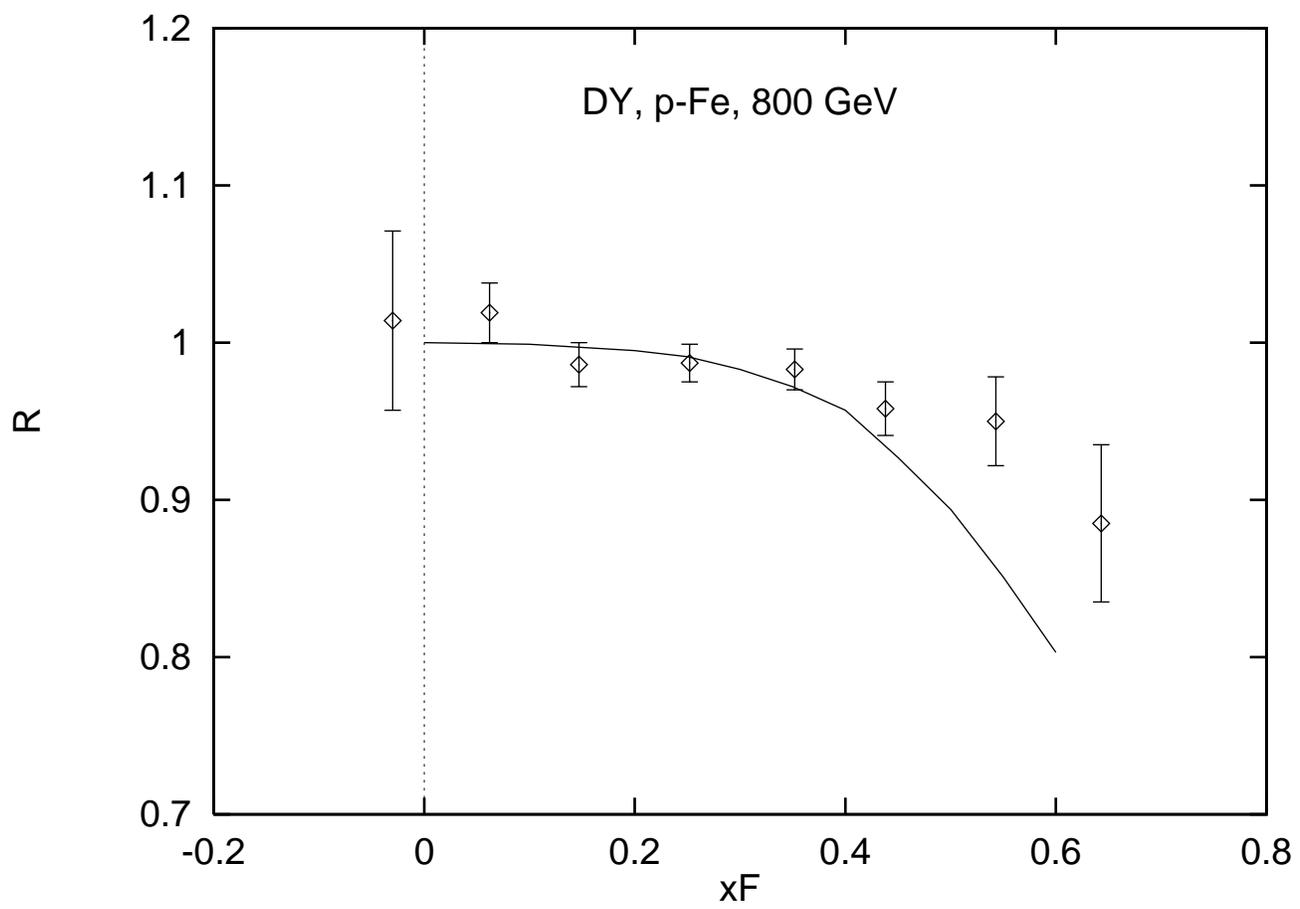
\end{document}